\documentclass[prl,aps,twocolumn,amsmath,amssymb,floatfix,pra,reprint,footinbib,superscriptaddress,showpacs,longbibliography]{revtex4-1}

\usepackage{lipsum}

\usepackage{txfonts}
\usepackage{mathrsfs,times}
\usepackage{ulem}
\usepackage{textcomp}

\usepackage{graphicx}
\usepackage{epstopdf}
\usepackage[applemac]{inputenc}
\usepackage[T1]{fontenc}
\usepackage[english]{babel}
\usepackage{ae}
\usepackage{siunitx}
\usepackage{color}
\usepackage{url}
\usepackage{amsmath,amssymb,natbib}
\usepackage{psfrag}
\usepackage{fixmath}
\usepackage{booktabs}
\usepackage{slashed}
\usepackage[americaninductors]{circuitikz}
\usepackage{tikz}
\usetikzlibrary{arrows}

\usepackage[colorlinks]{hyperref}
\hypersetup{%
        plainpages=true,
        breaklinks=true,
        hypertexnames=false,
        pageanchor=true,
        colorlinks=true,
        linkcolor={blue},
        citecolor={red},
        urlcolor={blue},
        anchorcolor={black}
      }
       
\usepackage{mleftright} 

\newcommand{\ket}[1]{|#1\rangle}
\newcommand{\be}{\begin{equation}}
\newcommand{\ee}{\end{equation}}
\newcommand{\bea}{\begin{eqnarray}}
\newcommand{\eea}{\end{eqnarray}}

\newcommand{\ketbra}[2]{\mleft| #1 \rangle \langle #2 \mright|}

\newcommand{\abs}[1]{\mleft|#1\mright|}
\newcommand{\abssq}[1]{\mleft| #1 \mright|^2}
\newcommand{\comm}[2]{\mleft[ #1, #2 \mright]}

\newcommand{\sm}{\sigma_-}
\renewcommand{\sp}{\sigma_+}
\newcommand{\nl}{\nonumber \\}
\newcommand{\nn}{\nonumber}
\newcommand{\figref}[1]{\mbox{Fig.~\ref{#1}}}

\renewcommand{\eqref}[1]{\mbox{Eq.~(\ref{#1})}}

\makeatletter
\newcommand*{\rom}[1]{\expandafter\@slowromancap\romannumeral #1@}
\makeatother

\begin{document}

\title {Oscillating bound states for a giant atom}

\author{Lingzhen Guo}
\affiliation{Max Planck Institute for the Science of Light, Staudtstra\ss{}e~2, 91058 Erlangen, Germany}

\author{Anton Frisk Kockum}
\affiliation{Department of Microtechnology and Nanoscience (MC2), Chalmers University of Technology, SE-41296 G\"oteborg, Sweden}

\author{Florian Marquardt}
\affiliation{Max Planck Institute for the Science of Light, Staudtstra\ss{}e~2, 91058 Erlangen, Germany}
\affiliation{Physics Department, University of Erlangen-Nuremberg, Staudtstra\ss{}e~5, 91058 Erlangen, Germany}

\author{G\"oran Johansson}
\affiliation{Department of Microtechnology and Nanoscience (MC2), Chalmers University of Technology, SE-41296 G\"oteborg, Sweden}

\date{\today}

\begin{abstract}
We investigate the relaxation dynamics of a single artificial atom interacting, via multiple coupling points, with a continuum of bosonic modes (photons or phonons) in a one-dimensional waveguide. In the non-Markovian regime, where the travelling time of a photon or phonon between the coupling points is sufficiently large compared to the inverse of the bare relaxation rate of the atom, we find that a boson can be trapped and form a stable bound state. More interestingly, if the number of coupling points is more than two, the bound state can oscillate persistently by exchanging energy with the atom despite the presence of the dissipative environment. We propose several realistic experimental schemes to generate such oscillating bound states. 
\end{abstract}


\maketitle

\begin{figure}
\includegraphics[width=0.99\linewidth]{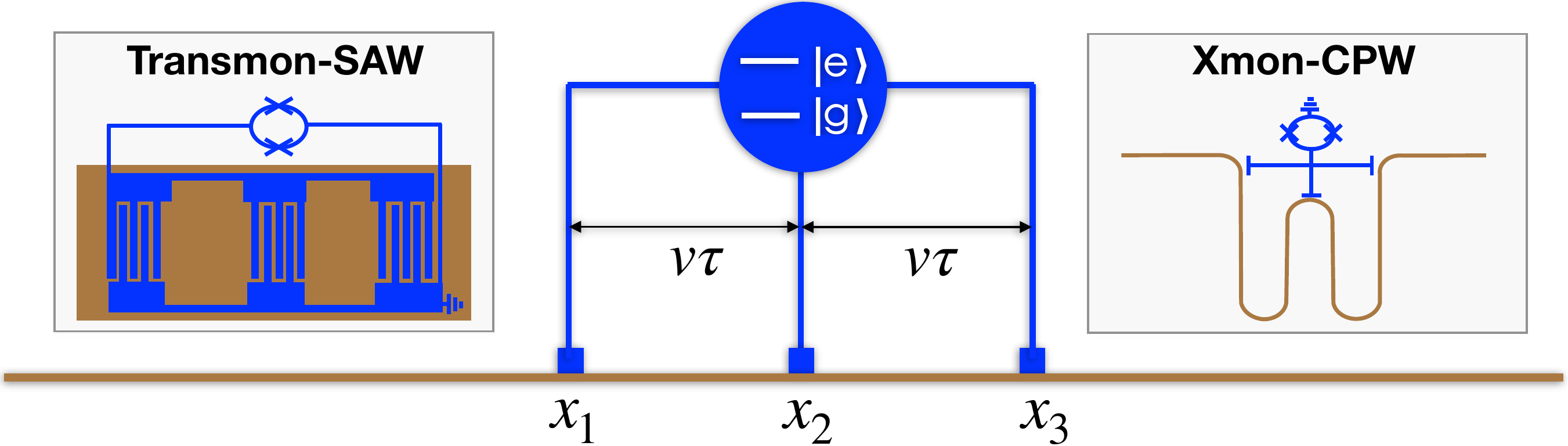}
\caption{Sketch and experimental setups for giant atoms. An atom (blue) couples to a waveguide (brown) at multiple points $x_j$, which are spaced far apart. Left: a transmon qubit coupled to a SAW waveguide via multiple interdigital transducers. Right: an Xmon qubit coupled capacitively to a meandering microwave CPW at multiple points.
\label{fig_ModelHamiltonian}}
\end{figure}

\textit{Introduction.---}%
The study of interaction between light and matter is one of the core topics in modern physics~\cite{Kockum2019}. In such studies, the wavelength of the light is usually large compared to the size of the (artificial) atoms constituting the matter~\cite{Goy1983, Leibfried2003, Wallraff2004, Miller2005, Haroche2013, Gu2017}. Indeed, the traditional framework of quantum optics is based on point-like atoms~\cite{Walls2008} and neglects the time it takes for light to pass a single atom. Recently, following significant technological advances for superconducting circuits~\cite{You2011, Gu2017, Kockum2019a, Krantz2019}, ``giant'' artificial atoms (transmon qubits~\cite{Koch2007}) have been designed to interact with surface acoustic waves (SAWs) via multiple coupling points in a waveguide~\cite{Gustafsson2014, Aref2016, Andersson2019} (or resonator~\cite{Manenti2017, Noguchi2017, Moores2018, Satzinger2018, Bolgar2018, Sletten2019, Bienfait2019}) as sketched in \figref{fig_ModelHamiltonian} (left inset). Such a giant-atom structure can also be realised in a more conventional circuit-quantum-electrodynamics (circuit-QED) experiment by coupling a single Xmon~\cite{Barends2013}, a version of the transmon, to a meandering coplanar waveguide (CPW) as sketched in \figref{fig_ModelHamiltonian} (right inset)~\cite{Kockum2014, Kannan2019, Vadiraj2019}. Since the distance between coupling points can be (much) longer than the characteristic wavelength of the bath, it is necessary to consider the phase difference between these coupling points. Striking effects have been found as a consequence of this, e.g., frequency-dependent relaxation rate and Lamb shift of a giant atom~\cite{Kockum2014, Kannan2019, Vadiraj2019}, and decoherence-free interaction between multiple giant atoms~\cite{Kockum2018, Kannan2019}. The giant-atom scheme has recently been extended to higher dimensions with cold atoms~\cite{Gonzalez-Tudela2019} and constitutes an exciting new paradigm in quantum optics~\cite{Gonzalez-Tudela2019, Kockum2019a}, where much remains to explore.

Spurred by the growing interest in quantum information science, there have been many investigations of non-Markovian open quantum systems, e.g., single atom(s) in front of a mirror~\cite{Eschner2001, Dorner2002, Tufarelli2014, Guimond2017, Calajo2019} or distant atoms coupled locally to the same environment~\cite{Milonni1974, Zheng2013a, Gonzalez-Ballestero2013, Laakso2014, Gonzalez-Ballestero2014, Fang2015, Guimond2016a, Ramos2016}. The physical origin of the non-Markovianity is typically the coupling to a structured bath causing information back-flow from the environment~\cite{Breuer2016, Rist2008, Gribben2019}. These systems can exhibit non-exponential relaxation~\cite{Tong2010, Garmon2013} and bound states~\cite{Tanaka2006, Zheng2013a, Calajo2016, Shi2016, Facchi2016, Gonzalez-Tudela2017a, Gonzalez-Tudela2017d, Liu2017, Sundaresan2019}, which can be harnessed for quantum simulations~\cite{Douglas2015, Gonzalez-Tudela2018}. Here, we realize non-Markovianity in a single giant atom by engineering the time delays between coupling points to be comparable to the relaxation time~\cite{Guo2017, Ask2019}. For such a non-Markovian giant atom with \textit{two} coupling points, it has been predicted~\cite{Guo2017}, and recently observed in experiment~\cite{Andersson2019}, that the spontaneous decay is polynomial instead of exponential. 

In this Letter, we investigate the relaxation dynamics of a single giant atom interacting with a one-dimensional (1D) bosonic bath (e.g., an open waveguide for phonons or photons) through \textit{multiple} coupling points. Our main result is that \textit{three or more} coupling points enable the creation of \textit{persistently oscillating bound states}, a phenomenon which, to the best of our knowledge, is unique to giant atoms. We envision that this phenomenon could be used in quantum information processing as a  single-photon (-phonon) ``tweezer" or trap, and that it could be viewed as a minimalistic implementation of cavity QED with the atom forming its own cavity.


\textit{Model Hamiltonian.---}%
We consider a two-level atom interacting with an open 1D waveguide at $N$ coupling points |\figref{fig_ModelHamiltonian} illustrates the case $N=3$]. As illustrated by the two insets in \figref{fig_ModelHamiltonian}, this system can be implemented in at least two different experimental schemes: a transmon qubit with multiple interdigital transducers (IDTs) coupled to SAWs through piezoelectric effects~\cite{Gustafsson2014, Aref2016, Manenti2017, Andersson2019} or an Xmon qubit~\cite{Barends2013, Kockum2014, Kannan2019, Vadiraj2019} with multiple arms capacitively coupled to a coplanar waveguide. The total Hamiltonian for the system is
\bea
\label{TotalH}
H &=& \hbar \Omega \sp \sm + \int_{- \infty}^{+ \infty} dk \ \hbar \omega_k a^\dag_k a_k \nl
&&+ \sum_{m = 1}^N \hbar \sqrt{\frac{\gamma v}{2 \pi}} \int_{- \infty}^\infty dk \ \mleft( e^{i k x_m} a_k \sp + \text{h.c.} \mright),
\eea
where we have defined the atomic operators $\sp = \ketbra{e}{g}$ and $\sm = (\sp)^\dag$ with $\ket{g}$ ($\ket{e}$) the atomic ground (excited) state and $\Omega$ the atomic transition frequency. The parameters $k$, $v$, and $\omega_k = \abs{k} v$ are the wave vectors, velocities, and frequencies of the bosonic fields (phonons or photons) in the waveguide. The field operators $a_k$ satisfy $\comm{a_k}{a^\dag_{k'}} = \delta (k - k')$. The rotating-wave approximation (RWA) has been applied in the interaction term. We assume a constant effective relaxation rate $\gamma$ at each coupling point, located at $x_m$ $(m = 1, 2, \cdots, N)$. We also assume the coupling points are equidistant. Thus, the travel time for bosons between two neighbouring coupling points is a constant $\tau = (x_{m+1} - x_m) / v$. In this work, we investigate novel phenomena arising from non-Markovian dynamics due to $\tau$ being non-negligible. 


\textit{Equations of motion and their solutions.---}%
We study the process of spontaneous emission from the giant atom into the waveguide. The atom begins in the excited state $\ket{e}$ and the field in the waveguide is in the vacuum state $\ket{vac}$. Since the total number of atomic and field excitations is conserved in \eqref{TotalH} due to the RWA, we study the single-excitation subspace of the full system. The total system state can thus be described by
\be
\label{singlephoton}
\ket{\Psi (t)} = \beta (t) \ket{e, vac} + \int dk \ \alpha_k (t) a^\dag_k \ket{g, vac},
\ee
where the integral describes the state of a single boson propagating in the waveguide. From the Schr\"odinger equation $i \hbar \partial/ \partial t \ket{\Psi (t)} = H \ket{\Psi (t)}$, following the method in Ref.~\cite{Guo2017}, we derive the equation of motion (EOM) for the probability amplitude of the giant atom being excited,
\be
\label{bteom}
\frac{d}{dt} \beta (t) = - i \Omega \beta (t) - \frac{1}{2} N \gamma \beta (t) - \gamma \sum_{l=1}^{N-1} (N - l) \beta (t - l\tau) \Theta (t - l \tau),
\ee
and the time evolution of the bosonic field function $\varphi (x, t) \equiv \int_{- \infty}^\infty dk \ e^{i (k x - \omega_k t)} \alpha_k (t)$ in the waveguide
\be
\label{totalfield}
\varphi (x, t) = - i \sqrt{\frac{\gamma}{2 v}}\sum^N_{m = 1} \beta \mleft( t - \frac{\abs{x - x_m}}{v} \mright) \Theta \mleft( t - \frac{\abs{x-x_m}}{v} \mright).
\ee
Here, $\Theta (\bullet)$ is the Heaviside step function, which describes time-delayed feedback among the coupling points. The field intensity function $p (x, t) \equiv \abssq{\varphi (x, t)}$ describes the probability density at position $x$ and time $t$ to find a single phonon or photon for all possible wave vectors $k$.

The first term on the right-hand side of \eqref{bteom} describes the coherent dynamics of the atom. The second and third terms describe the relaxation processes due to Markovian and non-Markovian dynamics, respectively. The solution of $\beta (t)$ can be obtained by a Laplace transformation:
\be
\label{tbeta}
\beta (t) = \sum_n \frac{e^{s_n t}}{1 - \gamma \tau \sum_{l = 1}^{N - 1} (N - l) l e^{- s_n l \tau}},
\ee
where the complex frequency parameters $s_n$ are given by the solutions to the equation
\be
\label{Nlegspoles}
s_n + i \Omega + \frac{1}{2} N \gamma + \gamma \sum_{l = 1}^{N - 1} (N - l) e^{- s_n l \tau} = 0.
\ee
%
For finite time delay $\tau >0$, the nonlinear \eqref{Nlegspoles} has multiple solutions. In general, there is no simple closed form for these solutions.


\textit{Dark-state condition.---}%
Usually, the complex frequency $s_n$ has a negative real part, which represents the relaxation rate. In some particular situations, $s_n$ can be purely imaginary. In that case, the corresponding mode is a \textit{dark state}, which does not decay despite the dissipative environment. We seek the purely imaginary solution $s_n \equiv - i \Omega_n$ with
\be
\label{Omegan}
\Omega_n = \frac{2 n \pi}{N \tau}, \quad n \in \mathbb{Z}.
\ee
Plugging this into \eqref{Nlegspoles}, we obtain the following condition for the dark states:
\be
\label{darkcondition}
\Omega \tau = \frac{2 n \pi}{N} - \frac{1}{2} N \gamma \tau \cot \mleft( \frac{n \pi}{N} \mright), \quad n \in \mathbb{Z}.
\ee
Note that for the RWA to hold, we require $\abs{\Omega_n - \Omega} / \Omega \ll 1$ or, equivalently, $\abs{\frac{N \gamma}{2 \Omega} \cot \mleft( \frac{n \pi}{N} \mright)} \ll 1$ and $n \in \mathbb{Z^+}$ according to \eqref{darkcondition}. In the Markov limit $\gamma \tau \rightarrow 0$, the dark-state condition \eqref{darkcondition} is simplified into $\Omega \tau = {2 n \pi} / N$ and the dark frequency is $\Omega_n = \Omega + \frac{1}{2} N \gamma \cot \mleft( \frac{n \pi}{N} \mright)$~\cite{Kockum2014}. In the non-Markovian limit of sufficiently large $\gamma \tau$, the additional nonlinear cotangent term in \eqref{darkcondition} cannot be neglected. Due to this term, there is an associated bound field state in the waveguide for a given dark state of the atom.


\textit{Bound states.---}%
Inserting the dark-state solution $s_n = - i \frac{2 n \pi}{N \tau}$ into \eqref{tbeta}, we obtain the long-time dynamics of the atomic excitation probability amplitude
\be
\label{darkevolution}
\beta (t) \rightarrow A (n) e^{- i \frac{2 n \pi}{N \tau} t} \ \ \mathrm{with} \ \ A (n) = \frac{2 \sin^2 \mleft( n \pi / N \mright)}{2 \sin^2 \mleft( n \pi / N \mright) + N \gamma \tau}.
\ee
From Eqs.~(\ref{totalfield}) and (\ref{darkevolution}), we calculate~\cite{SupMat} the explicit expression for the field density in the long-time limit, $p_n (x) \equiv p (x, t \rightarrow \infty)$, for a given dark state $s_n$:
\be
\label{ndensity}
p_n (x) = \frac{8 \gamma}{v} \frac{\sin^2 \frac{n \pi}{N} \sin^2 \mleft( \frac{n \pi}{N} m' \mright)}{\mleft( 2 \sin^2 \frac{n \pi}{N} + N \gamma \tau \mright)^2} \sin^2 \mleft[ \frac{n \pi}{N} \mleft( m' + 2 \lambda - 1 \mright) \mright].
\ee
Here, we have relabelled the position coordinate by $x = (m' - 1 + \lambda) v \tau$ with $m' = 1, 2, \ldots, N$ and $\lambda \in [0,1)$. Equation (\ref{ndensity}) is only valid for the position between the two outermost coupling points, i.e., $x \in [x_1, x_N]$ with $x_1 = 0$ and $x_N = (N - 1) v \tau$. Outside the giant atom, i.e., for $x\notin[x_1, x_N]$, the field intensity $p_n (x)$ is zero.

\begin{figure}
\centering
\includegraphics[width=\linewidth]{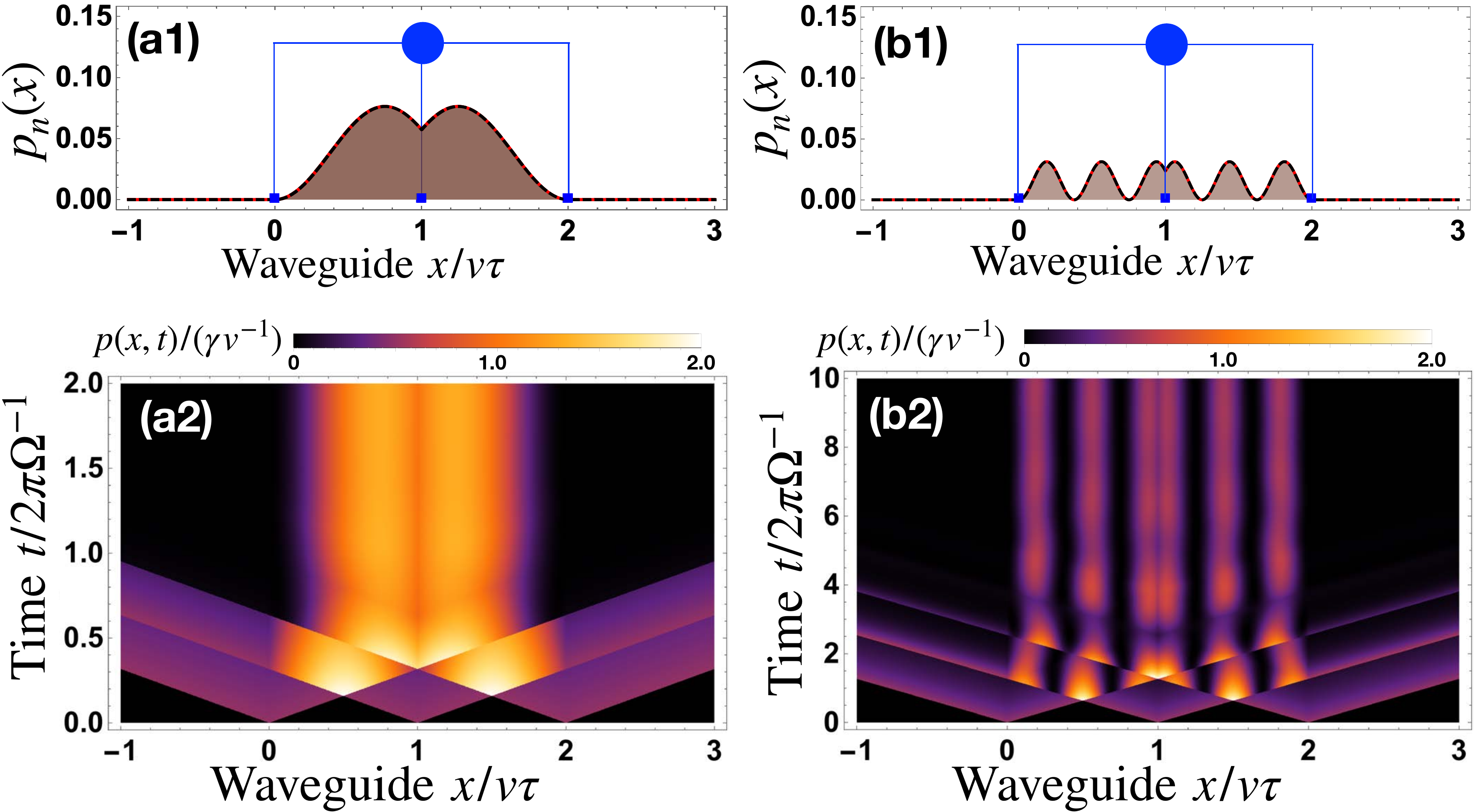}
\caption{Bound states in the waveguide for a giant atom with $N = 3$ coupling points. (a1) Field intensity distribution in the long-time limit and (a2) field intensity time evolution, for the dark state $s_{n=1}$ with $\gamma \tau / 2 \pi = 0.018$ and $\Omega \tau / 2 \pi = 0.317$. (b1, b2) Same, but for the dark state $s_{n=4}$ with $\gamma \tau / 2 \pi = 0.073$ and $\Omega \tau / 2 \pi = 1.27$. The red filled curves in (a1, b1) are numerical simulations and the black dashed lines are the analytical predictions from \eqref{ndensity}.
\label{Fig-Boundstates}}
\end{figure}

We calculate~\cite{SupMat} the total field intensity $I (n)$ of the bound field state for a given dark state:
\be
\label{p}
I (n) \equiv \int p_n (x) dx = \frac{2 N \gamma \tau \sin^2 \frac{n \pi}{N}}{\mleft( 2 \sin^2 \frac{n \pi}{N} + N \gamma \tau \mright)^2} \mleft(1 + \frac{N}{4 n \pi} \sin \frac{2 n \pi}{N} \mright).
\ee
We see that, in the Markovian limit $\gamma \tau \rightarrow 0$, the total field strength $I (n) \rightarrow 0$. Thus, the bound state only exists in the non-Markovian regime, where $\gamma\tau$ is sufficiently large. In the special case of $N = 2$, the dark-state condition \eqref{darkcondition} can only be fulfilled for odd integers $n$, and the residual field strength is $I (n) = \gamma \tau / (1 + \gamma \tau)^2 \leq 1/4$. In \figref{Fig-Boundstates}, we show how the bound state is formed. We plot the long-time field intensity distribution $p_n (x)$ [Figs.~\ref{Fig-Boundstates}(a1) and (b1)] and the time evolution of the field intensity function $p (x, t)$ [Figs.~\ref{Fig-Boundstates}(a2) and (b2)] for two different dark states ($n = 1$ and $n = 4$) of a giant atom with $N = 3$ coupling points.


\begin{figure}
\includegraphics[width=\linewidth]{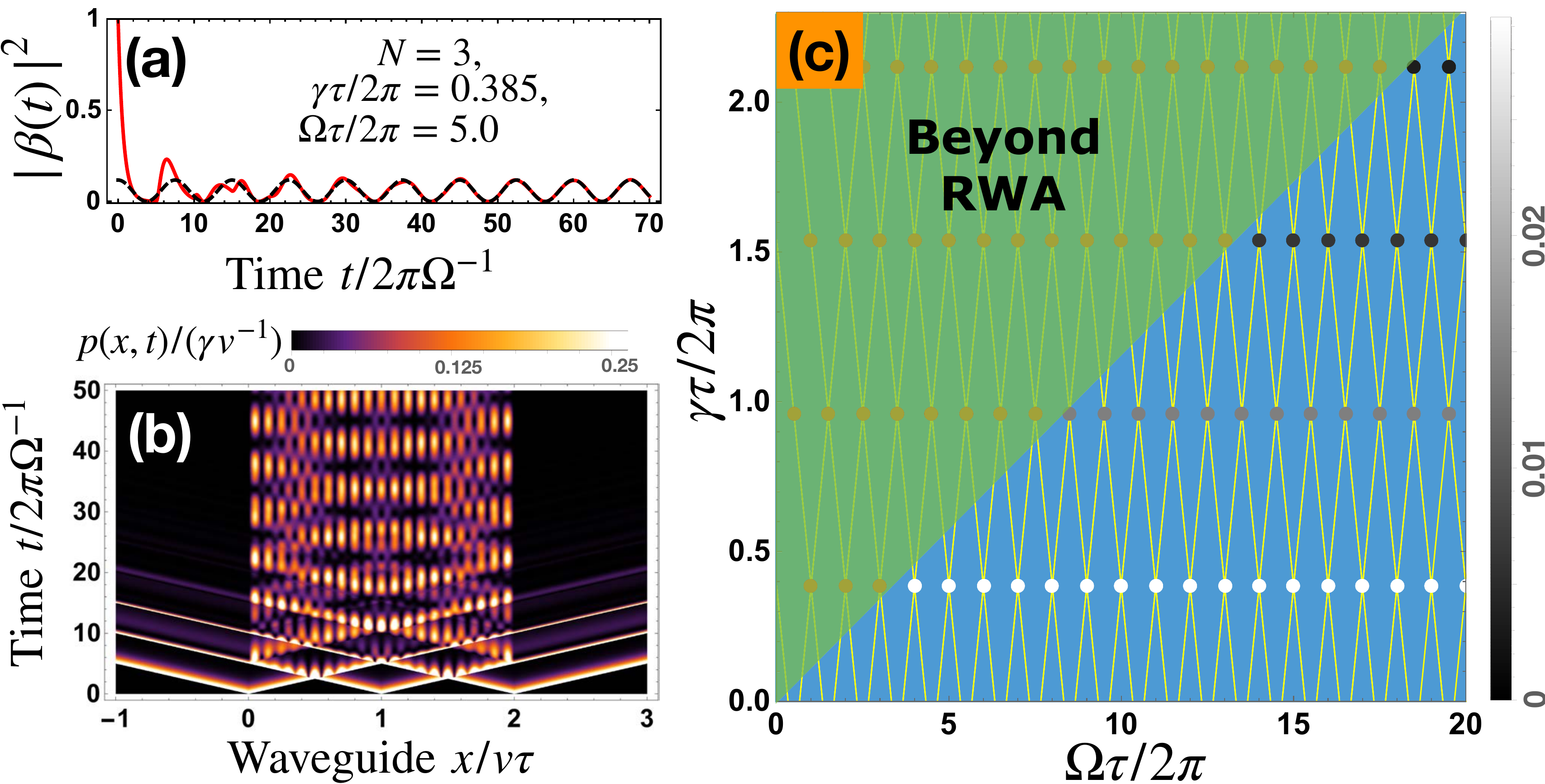}
\caption{Oscillating bound states for a giant atom with $N=3$. (a) Time evolution of the atomic excitation probability $\abssq{\beta (t)}$ with two coexisting dark states $s_{n = 14}$ and $s_{n = 16}$, from the numerical simulation (red solid line) and the analytical result (black dashed line) of \eqref{bidarkstateenergy}. (b) Time evolution of the field intensity $p (x, t)$ in the waveguide with the same parameters as in panel (a). (c) Conditions for oscillating bound states (solid dots) in the $\Omega \tau$ - $\gamma \tau$ parameter plane. The dots in the green region are beyond the RWA. The gray colour level of the dots in the RWA region indicates the oscillating amplitude of $A (n_1) A (n_2)$. The yellow lines show the conditions for non-oscillating bound states [as in \figref{Fig-Boundstates}] from \eqref{darkcondition} with fixed integers $n \in \mathbb{Z^+}$.
\label{Fig-OscillatingDarkSates}}
\end{figure}

\textit{Oscillating bound states.---}%
The dark-state condition \eqref{darkcondition} is a nonlinear equation for integer $n$ and $\gamma \tau > 0$. It is possible to find two integers $n_1$ and $n_2$ satisfying \eqref{darkcondition} simultaneously. This means that, in the long-time limit after all the dissipative modes die out, the dynamics of the atomic excitation probability amplitude $\beta(t)$ is a superposition of two dark states with different frequencies $\Omega_{n_1}$ and $\Omega_{n_2}$. As a result, the atomic excitation probability $\abssq{\beta (t)}$ oscillates persistently with frequency $\Omega_{n_1} - \Omega_{n_2}$ despite the dissipative environment. In \figref{Fig-OscillatingDarkSates}(a), we show the population dynamics for a three-leg giant atom ($N = 3$) with two coexisting dark states: $s_{n = 14}$ and $s_{n = 16}$. The undamped oscillation of $\abssq{\beta (t)}$ indicates that the atom exchanges energy with the bosonic bath persistently. In \figref{Fig-OscillatingDarkSates}(b), we plot the corresponding time evolution of the field intensity in the waveguide, showing an oscillating bound state in the long-time limit. In the Supplemental Video, we show an animation of the time evolution for the atomic excitation probabilitiy and the field intensity in the waveguide.

If $n_1$ and $n_2$ are the two simultaneous solutions of \eqref{darkcondition}, the parameters $\Omega \tau$ and $\gamma \tau$ have to be
%
\be
\label{bidarks3}
\mleft\{ \begin{array}{l}
\Omega \tau = \frac{2 n_1 \pi}{N} - \frac{2 (n_1 - n_2) \pi}{N} \frac{\cot\mleft( \frac{n_1 \pi}{N} \mright)}{\cot \mleft( \frac{n_1 \pi}{N} \mright) - \cot \mleft( \frac{n_2 \pi}{N} \mright)} > 0,
\\
\gamma \tau = \frac{4 (n_1 - n_2) \pi}{N^2} \frac{1}{\cot \mleft( \frac{n_1 \pi}{N} \mright) - \cot \mleft( \frac{n_2 \pi}{N} \mright)} > 0.
\end{array} \mright.
\ee
Here, the physical conditions of $\Omega \tau > 0$ and $\gamma \tau > 0$ need to be satisfied, together with the RWA condition that $\abs{\frac{N \gamma}{2 \Omega} \cot \mleft( \frac{n_{1(2)} \pi}{N} \mright)} \ll 1$ and $n_{1(2)} \in \mathbb{Z^+}$. The long-time dynamics of the giant atom is $\beta (t) \rightarrow A (n_1) e^{- i \Omega_{n_1} t} + A (n_2) e^{- i \Omega_{k_2} t}$, which results in
%
\be
\label{bidarkstateenergy}
\abssq{\beta (t)} = A^2 (n_1) + A^2 (n_2) + 2 A (n_1) A (n_2) \cos \mleft[ \mleft( \Omega_{n_1} - \Omega_{n_2} \mright) t \mright].
\ee
The amplitude of the persistent oscillations is thus $A (n_1) A (n_2)$. 

The total field intensity left in the waveguide for two coexisting dark states is $I (n_1, n_2) \equiv \int p (x, t \rightarrow \infty) dx$, which can be calculated from Eqs.~(\ref{totalfield}) and (\ref{tbeta})~\cite{SupMat}:
\be
\label{PTbidark2}
I (n_1, n_2) = I (n_1) + I (n_2) - 4 A (n_1) A (n_2) \frac{\Omega \cos \mleft[ \mleft( \Omega_{n_1} - \Omega_{n_2} \mright) t \mright]}{\Omega_{n_1} + \Omega_{n_2}}.
\ee
%
According to \eqref{singlephoton}, the quantity $\abssq{\beta (t)} + I (n_1, n_2)$ is the total excitation probability of the atom and the field, which is conserved, since the oscillating bound state does not decay. This gives an additional condition for the coexisting dark states:
\be
\frac{\Omega_{n_1} + \Omega_{n_2}}{2} = \Omega.
\ee
Combing this with \eqref{darkcondition}, we find that the solutions are of the form $n_1 = p N + n$ and $n_2 = q N - n$ with $p, q \in \mathbb{Z^+}$ and $1 \leq n < N$. The conditions in \eqref{bidarks3} then become $\Omega \tau / 2 \pi = (p + q) / 2$ and $\gamma \tau / 2 \pi = \mleft[ (p - q) / N + 2 n / N^2 \mright] \tan \mleft( \frac{n \pi}{N} \mright)$. By setting $p \geq q$ and $1 \leq n < N / 2$, \eqref{bidarks3} can be satisfied and we obtain the frequencies of the two dark modes: $\Omega \pm \frac{1}{2} N \gamma \cot \mleft( \frac{n \pi}{N} \mright)$.

In \figref{Fig-OscillatingDarkSates}(c), we show the existence of oscillating bound states (solid dots) in the $\Omega \tau$ - $\gamma \tau$ parameter space for a giant atom with $N = 3$. The condition in \eqref{bidarks3} implies that, if $n_1$ and $n_2$ are solutions yielding coexisting dark states, the integers $n_1 + N$ and $n_2 + N$ are also solutions of coexisting dark states with $\gamma \tau$ unchanged but $\Omega \tau$ increased by $2 \pi$. This results in the $2 \pi$ periodicity along the horizontal direction in \figref{Fig-OscillatingDarkSates}(c). The dots in the green region are beyond RWA, where the dark-mode frequency $\abs{\Omega_{n_{1(2)}} - \Omega} / \Omega > 0.1$. 
%

If the giant atom only has two coupling points ($N = 2$), the nonlinear cotangent term in condition (\ref{darkcondition}) is either zero or infinity. Therefore, \textit{the oscillating bound states only exist for more than two coupling points ($N \geq 3$)}.


\textit{Continuum limit.---}%
We now discuss the limit of infinitely many coupling points ($N \rightarrow \infty$). In this case, the time it takes for the field in the waveguide to pass all coupling points is $N \tau \rightarrow T$. For capacitive coupling between the atom and the waveguide, the interaction strength $g$ at a single point is proportional to the local capacitance $c$, i.e., $g \propto c$~\cite{Koch2007, Andersson2019} and the relaxation rate is $\gamma \propto g^2 \propto c^2$~\cite{Kockum2014, Guo2017}. As a result, the parameter $N^2 \gamma \propto (N c)^2$, where $N c$ is the total capacitance, is a converged quantity $N^2 \gamma \rightarrow \Gamma$, which describes the total relaxation rate of the atom into the waveguide. In this continuum limit, the dark-state condition \eqref{darkcondition} becomes
\be
\label{darkcondition2}
\Omega T = 2 n \pi - \frac{\Gamma T}{2 n \pi}, \quad n \in \mathbb{Z}.
\ee
The solution is $n = \mleft( 4 \pi \mright)^{-1} \mleft[ \Omega T \pm \sqrt{(\Omega T)^2 + 4 \Gamma T} \mright] \in \mathbb{Z}$, and the corresponding dark-mode frequency is $\Omega_n = \Omega + \frac{\Gamma }{2 n \pi}$. The field intensity $p_n (x)$ of the bound state can be calculated from \eqref{ndensity}, yielding
\be
\label{ndensitycontinuum}
p_n (x) = \frac{2 n^2 \pi^2 / \Gamma T}{\mleft( 2 n^2 \pi^2 / \Gamma T + 1 \mright)^2} \frac{4}{L} \sin^4 \mleft( \frac{n \pi}{L} x \mright),
\ee
where $L = x_N - x_1$. The total field intensity of the bound state is $I (n) = \mleft( 3 n^2 \pi^2 / \Gamma T \mright) \mleft( 2 n^2 \pi^2 / \Gamma T + 1 \mright)^{-2} \leq 3/8$. However, since the RWA condition requires $n>0$, and we only have one solution fulfilling that condition, it is not expected that an oscillating bound state can be created in this case.
%

\begin{figure}
\centering
\includegraphics[width=\linewidth]{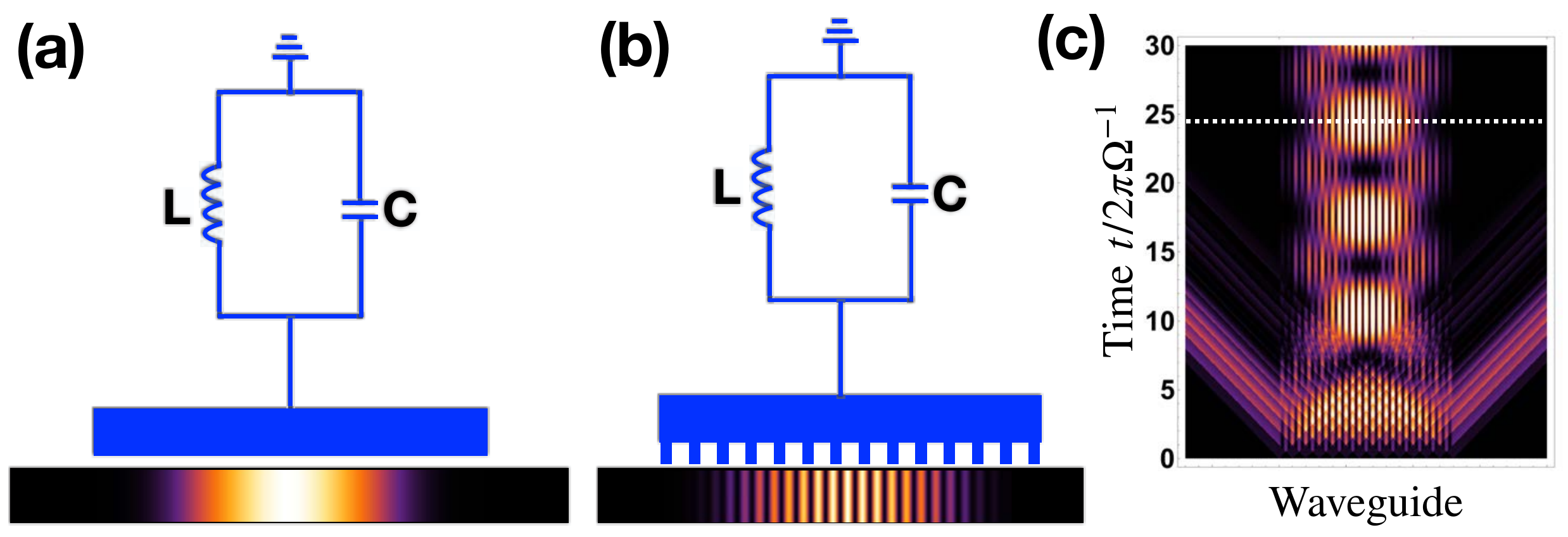}
\caption{Experimental setups for generating (a) a static bound state with a continuum metal and (b) an oscillating bound state with a comb-like metal.  In both cases, the metals are coupled to a 1D SAW waveguide. The colours in the waveguides show the field intensity of the bound states. For the oscillating bound state in (b), the field intensity is taken at the fixed moment indicated by the white dashed line on the plot of $p (x, t)$ in (c). The attached LC circuits are used to tune the plasmon frequency $\Omega$ in the metal. Parameters: (a) $n = 1$; (b) and (c) $\Omega \tau = 2 \pi$, $\Gamma T \rightarrow 4 \pi^2$ (i.e., $n = 1$).
\label{Fig-LC}}
\end{figure}

Note that the EOM (\ref{bteom}) also describes the linear (classical) problem where a single harmonic mode instead of an atom interacts with the continuum of modes in an infinite waveguide. Therefore, our predictions can be immediately applied to this linear (classical) system. In \figref{Fig-LC}(a), we show a continuum metal contacting capacitively with an infinite SAW waveguide made of piezoelectric material. The metal is attached to an $LC$ circuit to tune the plasmon frequency in the metal. If the dark condition in \eqref{darkcondition2} is satisfied, we expect to observe a bound state in the waveguide. To generate an oscillating bound state, we can design the contact part of the metal as a comb-like structure as shown in \figref{Fig-LC}(b). Note that the two integers $n_1 = N + n$ and $n_2 = N - n$ with $1 \leq n < N / 2$ always satisfy the dark-state condition \eqref{bidarks3}. In the limit of infinitely many coupling points $N \gg n$, i.e., for a very extended comb, we have $\Omega \tau = 2 \pi$ and $\Gamma T \rightarrow (2 n \pi)^2$. In this parameter setting, we can create two coexisting dark modes with frequencies $\Omega_{\pm} \rightarrow \Omega \pm \frac{\Gamma}{2 \pi n}$. We show the field intensity of bound states in the 1D waveguide for the dark state $n = 1$ in \figref{Fig-LC}(b) and (c).


\textit{Discussion and conclusion.---}%
%
We have shown that a giant atom with $N \geq 3$ coupling points to an open waveguide can harbour \textit{oscillating bound states}. To observe these states in experiment, the coherence time of the (artificial) atom must exceed the oscillation period. For a transmon or Xmon qubit, the coherence time can be on the order of hundreds of microseconds~\cite{Barends2013, Oliver2013, Gu2017, Burnett2019, Kjaergaard2019}, which is much longer than the oscillation period shown in \figref{Fig-OscillatingDarkSates}(a) since, typically, $\Omega / 2 \pi$ is several gigahertz. 
%
%

One application of these bound states in quantum information processing could be as a single-photon (-phonon) trap. Furthermore, the oscillating bound state, i.e., the dynamical exchange of excitations between the atom and the bosonic bound state, indicates that it is possible to realise a minimal version of cavity QED with a single giant atom in the open waveguide, reminiscent of the recent demonstration of cavity QED with atom-like mirrors~\cite{Mirhosseini2019}.


\begin{acknowledgments}

\textit{Acknowledgments.---}%
AFK acknowledges support from the Swedish Research Council (Grant No.~2019-03696). AFK and GJ acknowledge support from the Knut and Alice Wallenberg Foundation.

\end{acknowledgments}

\bibliography{GiantAtomRefsArXiv}



\onecolumngrid

\section{Supplemental Material for \\ \textit{Oscillating bound states for a giant atom} }

\subsection{I. Field intensity distribution for a single bound state}
In this section, we derive Eqs.~(10) and (11) in the main text. For a given dark mode $s_n = - i\frac{2 n \pi}{N \tau}$, the corresponding field intensity can also be calculated from Eqs.~(4) and (9) in the main text. By parametrizing the position coordinate as $x = (m' - 1) v \tau + \lambda v \tau$ with $m' = 1, 2, \ldots, N$  and $\lambda \in [0,1)$, we have $p_n (x) = p \mleft( x, t \rightarrow + \infty \mright)$ and
\bea
\label{densityp}
p \mleft( x, t \rightarrow + \infty \mright) &=& \frac{\gamma}{2 v} \abssq{\sum_m \beta \mleft( t - \abs{x - x_m} / v \mright) \Theta \mleft( t - \abs{x - x_m} / v \mright)} \nn\\
&=& \frac{\gamma}{2 v} \mleft( \frac{1}{1 + \frac{1}{2} \frac{N \gamma \tau}{\sin^2 (n \pi / N)}} \mright)^2 
\abssq{\sum_m \exp \mleft[i \frac{2 n \pi}{N \tau} \frac{\abs{x - x_m}}{v} \mright]} \nn\\
&=& \frac{\gamma}{2 v} \mleft( \frac{1}{1 + \frac{1}{2} \frac{N \gamma \tau}{\sin^2 (n \pi / N)}} \mright)^2 
\frac{1}{4} \abssq{\frac{4}{1 - e^{- i \frac{2 n \pi}{N}}}} 
\mleft[ 1 - \cos \mleft( \frac{2 n \pi}{N} m' \mright) \mright] 
\mleft[ 1 - \cos \mleft( \frac{2 n \pi}{N} \mleft[ m' + 2 \mleft( \lambda - \frac{1}{2} \mright) \mright] \mright) \mright] \nn\\
&=& \frac{\gamma}{2 v \sin^2 (n \pi / N)} \mleft( \frac{1}{1 + \frac{1}{2} \frac{N \gamma \tau}{\sin^2 (n \pi / N)}} \mright)^2
\mleft[ 1 - \cos \mleft( \frac{2 n \pi}{N} m' \mright) \mright]
\mleft[ 1 - \cos \mleft( \frac{2 n \pi}{N} \mleft[ m' + 2 \mleft( \lambda - \frac{1}{2} \mright) \mright] \mright) \mright] \nn\\
&=& \frac{2 \gamma}{v} \frac{\sin^2 \frac{n \pi}{N}}{\mleft( 2 \sin^2 \frac{n \pi}{N} + N \gamma \tau \mright)^2}
\mleft[ 1 - \cos \mleft( \frac{2 n \pi}{N} m' \mright) \mright]
\mleft[ 1 - \cos \mleft( \frac{2 k \pi}{N} \mleft[ m' + 2 \lambda - 1 \mright] \mright) \mright] \nn\\
&=& \frac{8 \gamma}{v} \frac{\sin^2 \frac{n \pi}{N}}{\mleft( 2 \sin^2 \frac{n \pi}{N} + N \gamma \tau \mright)^2} 
\sin^2 \mleft( \frac{n \pi}{N} m' \mright) \sin^2 \mleft[ \frac{k \pi}{N} \mleft( m' + 2 \lambda - 1 \mright) \mright].
\eea
This distribution is valid for $x$ between $x_1$ and $x_N$ in the waveguide. We see that at the two ends of the giant atom, $x_1 = 0$ (i.e., $m' = 1$ and $\lambda = 0$) and $x_N = (N - 1) v \tau$ (i.e., $m' = N$ and $\lambda = 0$), the intensity vanishes. When the position $x$ is outside the interval $[x_1, x_m]$, since the sign of $(x-x_m)$ is fixed, the summation in the second line gives zero. This is reasonable since the excitations outside the outermost coupling points will propagate away in the waveguide and never come back.

The total field intensity left in the waveguide for the given dark state $s_n = - i \frac{2 n \pi}{N \tau}$ can be calculated, in the long-time limit, by plugging $\beta (t)$ into the above equation, yielding
\be
\label{PT}
I(n) = \int_{x_1}^{x_N} p_n (x) \ dx = \frac{\gamma}{2} \mleft( \frac{1}{1 + \frac{1}{2} \frac{N \gamma \tau}{\sin^2 (n \pi / N)}}\mright)^2 \int_0^{\mathcal{T}} \abssq{\sum_m \exp \mleft( i \frac{2 n \pi}{N \tau} \abs{t' - \tau_m} \mright)} \ dt'.
\ee
Here, we have expressed the coordinates in terms of times, i.e., $t' \equiv x / v$ and $\tau_m = (m - 1) \tau$ with $m = 1, 2, \ldots, N$. The parameter $\mathcal{T} \equiv (N - 1) \tau$ is the total travelling time from $x_1$ to $x_N$. By parametrizing $t' = (m' - 1) \tau + a \tau$ with $a \in [0,1)$, we have
\bea
\label{integral}
\int_0^{\mathcal{T}}  \abssq{\sum_m e^{i \frac{2 n \pi}{N \tau} \abs{t' - \tau_m}}} \ dt '&=& \tau \sum_{m' = 1}^N \int_0^1 \abssq{\sum_{m = 1}^{m'} e^{i \frac{2 n \pi}{N \tau} \mleft[ (m' - m) \tau + a \tau \mright]} + \sum_{m = m' + 1}^N e^{i \frac{2 n \pi}{N \tau} \mleft[ (m - m') \tau - a \tau \mright]}} \ da \nn\\
&=& \tau \sum_{m' = 1}^N \int_0^1 \abssq{\frac{e^{i \frac{2 n \pi}{N} (m' + a - 1)} - e^{- i \frac{2 n \pi}{N}} e^{i \frac{2 n \pi}{N} a}}{1 - e^{- i \frac{2 n \pi}{N}}} + \frac{e^{i \frac{2 n \pi}{N} (1 - a)} - e^{i \frac{2 n \pi}{N}} e^{i \frac{2 n \pi}{N} (N - m' - a)}}{1 - e^{i \frac{2 n \pi}{N}}}} \ da \nn\\
&=& \tau \sum_{m' = 1}^N \int_0^1 \abssq{\frac{e^{i \frac{2 n \pi}{N} (m' + a - 1)} - e^{i \frac{2 n \pi}{N} (a - 1)}}{1 - e^{- i \frac{2 n \pi}{N}}} - \frac{e^{- i \frac{2 n \pi}{N} a} - e^{i \frac{2 n \pi}{N} (N - m' - a)}}{1 - e^{- i \frac{2 n \pi}{N}}}} \ da \nn\\
&=& \tau \abssq{\frac{1}{1 - e^{- i \frac{2 n \pi}{N}}}} \sum_{m' = 1}^N \int_0^1 \abssq{e^{i \frac{2 n \pi}{N} (a - 1)} \mleft( e^{i \frac{2 n \pi}{N} m'} - 1 \mright) + e^{- i \frac{2 n \pi}{N} a} \mleft( e^{- i \frac{2 n \pi}{N} m'} - 1 \mright)} \ da \nn\\
&=& \tau \abssq{ \frac{1}{1 - e^{- i \frac{2 n \pi}{N}}}} \sum_{m' = 1}^N \int_{- 1 / 2}^{1 / 2} \abssq{e^{i \frac{2 n \pi}{N} (a - 1  / 2)} \mleft( e^{ i \frac{2 n \pi}{N} m'} - 1 \mright) + e^{- i \frac{2 n \pi}{N} (a + 1 / 2)} \mleft( e^{- i \frac{2 n \pi}{N} m'} - 1 \mright)} \ da \nn\\
&=& \tau \abssq{\frac{1}{1 - e^{- i \frac{2 n \pi}{N}}}} \sum_{m' = 1}^N \int_{- 1 / 2}^{1 / 2} \abssq{e^{i \frac{2 n \pi}{N} a} \mleft(e^{i \frac{2 n \pi}{N} m'} - 1 \mright) + e^{- i \frac{2 n \pi}{N} a} \mleft( e^{- i \frac{2 n \pi}{N} m'}  - 1 \mright)} \ da \nn\\
&=& \tau \abssq{\frac{2}{1 - e^{- i \frac{2 n \pi}{N}}}} \sum_{m' = 1}^N \int_{- 1 / 2}^{1 / 2} \abssq{\cos \mleft( \frac{2 n \pi}{N} [m'+a] \mright) - \cos \mleft( \frac{2 n \pi}{N} a \mright)} \ da \nn\\
&=& \tau \abssq{\frac{4}{1 - e^{- i \frac{2 n \pi}{N}}}} \sum_{m' = 1}^N \abssq{\sin \mleft( \frac{n \pi}{N} m' \mright)} \int_{- 1  / 2}^{1 / 2} \abssq{\sin \mleft( \frac{n \pi}{N} [m' + 2 a] \mright)} \ da \nn\\
&=& \tau \frac{1}{2} \abssq{\frac{4}{1 - e^{- i \frac{2 n \pi}{N}}}} \sum_{m' = 1}^N \abssq{\sin \mleft( \frac{n \pi}{N} m' \mright)} \int_{- 1 / 2}^{1 / 2} \mleft[ 1 - \cos \mleft( \frac{2 n \pi}{N} [m' + 2 a] \mright) \mright] \ da \nn\\
&=& \tau \frac{1}{2} \abssq{\frac{4}{1 - e^{- i \frac{2 n \pi}{N}}}} \sum_{m' = 1}^N \sin^2 \mleft( \frac{n \pi}{N} m' \mright) \mleft[ 1 - \frac{N}{4 n \pi} \mleft( \sin \mleft[ \frac{2 n \pi}{N} (m'+1) \mright] - \sin \mleft[ \frac{2 n \pi}{N} (m' - 1) \mright] \mright) \mright] \nn\\
&=& \tau \frac{1}{4} \abssq{\frac{4}{1 - e^{- i \frac{2 n \pi}{N}}}} \sum_{m' = 1}^N \mleft[ 1 - \cos \mleft( \frac{2 n \pi}{N} m' \mright) \mright] \mleft[ 1 - \frac{N}{2 n \pi} \cos \mleft( \frac{2 n \pi}{N} m' \mright) \sin \mleft( \frac{2 n \pi}{N} \mright) \mright].
\eea
Using the identity $\sum_{m' = 1}^N \cos \mleft( \frac{2 n \pi}{N} m' \mright) = 0$, we obtain
\bea
\int_0^{\mathcal{T}} \abssq{\sum_m e^{i \frac{2 n \pi}{N \tau} \abs{t' - \tau_m}}} \ dt' &=&
\tau \frac{1}{4} \abssq{\frac{4}{1 - e^{- i \frac{2 n \pi}{N}}}} \sum_{m' = 1}^N \mleft[ 1 - \cos \mleft( \frac{2 n \pi}{N} m' \mright) \mright] \mleft[1 - \frac{N}{2 n \pi} \cos \mleft( \frac{2 n \pi}{N} m' \mright) \sin \mleft( \frac{2 n \pi}{N} \mright) \mright] \nn\\
&=& \tau \frac{1}{4} \abssq{\frac{4}{1 - e^{- i \frac{2 n \pi}{N}}}} \sum_{m' = 1}^N \mleft[ 1 + \frac{N}{2 n \pi} \cos^2 \mleft( \frac{2 n \pi}{N} m' \mright) \sin \mleft( \frac{2 n \pi}{N} \mright) \mright] \nn\\
&=& \tau \frac{1}{4} \abssq{\frac{4}{1 - e^{- i \frac{2 n \pi}{N}}}} \sum_{m' = 1}^N \mleft[ 1 + \frac{N}{4 n \pi} \sin \mleft( \frac{2 n \pi}{N} \mright) + \frac{N}{4 n \pi} \sin \mleft( \frac{2n\pi}{N} \mright) \cos \mleft( \frac{4 n \pi}{N} m' \mright) \mright] \nn\\
&=& \tau \abssq{\frac{1}{1 - e^{- i \frac{2 n \pi}{N}}}} 4 N \mleft[ 1 + \frac{N}{4 n \pi} \sin \mleft( \frac{2 n \pi}{N} \mright) \mright] = \frac{N \tau}{\sin^2 \mleft( \frac{n \pi}{N} \mright)} \mleft[ 1 + \frac{N}{4 n \pi} \sin \mleft( \frac{2 n \pi}{N} \mright) \mright] \nn\\
&=& 2 N \tau \frac{1 + \frac{N}{4 n \pi} \sin \mleft( \frac{2 n \pi}{N} \mright)}{1 - \cos \mleft( \frac{2 n \pi}{N} \mright)}.
\eea
Plugging the above result into Eq.~(\ref{PT}), we find the total field intensity left in the waveguide:
\be
\label{In}
I (n) = \frac{2 N \gamma \tau \sin^2 \mleft( \frac{n \pi}{N} \mright)}{\mleft[ 2 \sin^2 \mleft( \frac{n \pi}{N} + N \gamma \tau \mright) \mright]^2} \mleft[ 1 + \frac{N}{4 n \pi} \sin \mleft( \frac{2 n \pi}{N} \mright) \mright].
\ee


\subsection{II. Total field intensity of an oscillating bound state}

The total field intensity of an oscillating bound state with two coexisting dark states $s_{n_1}$ and $s_{n_2}$ is
\bea
\label{PTbidark1}
I (n_1, n_2) &=& \frac{\gamma}{2} \int_0^{\mathcal{T}} \abssq{ \sum_m A (n_1) e^{- i \frac{2 n_1 \pi}{N \tau} (t - \abs{t' - \tau_m})} + A (n_2) e^{- i \frac{2 n_2 \pi}{N \tau} (t - \abs{t' - \tau_m})}} \ dt' \nn\\
&=& \frac{\gamma \tau}{2} \sum_{m' = 1}^N \int_0^1
\mleft| A (n_1) e^{- i \frac{2 n_1 \pi}{N \tau} t} \mleft( \sum_{m = 1}^{m'} e^{i \frac{2 n_1 \pi}{N \tau} [(m' - m) \tau + a \tau]}
+ \sum_{m = m' + 1}^N e^{i \frac{2 n_1 \pi}{N \tau}[(m - m') \tau - a \tau]} \mright) \mright. \nn\\
&& \mleft. \qquad \qquad \qquad \qquad
+ A (n_2) e^{- i \frac{2 n_2 \pi}{N \tau} t} \mleft( \sum_{m = 1}^{m'} e^{i \frac{2 n_2 \pi}{N \tau}[(m' - m) \tau + a \tau]}
+ \sum_{m = m' + 1}^N e^{i \frac{2 n_2 \pi}{N \tau} [(m - m') \tau - a \tau]} \mright) \mright|^2 \ da \nn\\
&=& \frac{\gamma \tau}{2} \sum_{m' = 1}^N \int_0^1
\mleft| A (n_1) e^{- i \frac{2 n_1 \pi}{N \tau} t} \mleft( \frac{e^{i \frac{2 n_1 \pi}{N} (m' + a - 1)} - e^{i \frac{2 n_1 \pi}{N} (a - 1)}}{1 - e^{- i \frac{2 n_1 \pi}{N}}} 
- \frac{e^{- i \frac{2 n_1 \pi}{N} a} - e^{i \frac{2 n_1 \pi}{N}(N - m' - a)}}{1 - e^{- i \frac{2 n_1 \pi}{N}}} \mright) \mright. \nn\\
&& \mleft. \qquad \qquad \qquad \qquad
+ A (n_2) e^{- i \frac{2 n_2 \pi}{N \tau} t} \mleft( \frac{e^{i \frac{2 n_2 \pi}{N} (m' + a - 1)} - e^{i \frac{2 n_2 \pi}{N} (a - 1)}}{1 - e^{- i \frac{2 n_2 \pi}{N}}}
- \frac{e^{- i \frac{2 n_2 \pi}{N} a} - e^{i \frac{2 n_2 \pi}{N}(N - m' - a)}}{1 - e^{- i \frac{2 n_2 \pi}{N}}} \mright) \mright|^2 \ da \nn\\
&=& \frac{\gamma \tau}{2} \sum_{m' = 1}^N \int_0^1 \mleft|
\frac{A (n_1) e^{- i \frac{2 n_1 \pi}{N \tau} t}}{1 - e^{- i \frac{2 n_1 \pi}{N}}} \mleft[ e^{i \frac{2 n_1 \pi}{N} (a - 1)} \mleft( e^{i \frac{2 n_1 \pi}{N} m'} - 1 \mright) + e^{- i \frac{2 n_1 \pi}{N} a} \mleft( e^{- i \frac{2 n_1 \pi}{N} m'} - 1 \mright) \mright] \mright. \nn\\
&& \mleft. \qquad \qquad \qquad \qquad
+ \frac{A (n_2) e^{- i \frac{2 n_2 \pi}{N \tau} t}}{1 - e^{- i \frac{2 n_2 \pi}{N}}} \mleft[ e^{i \frac{2 n_2 \pi}{N} (a - 1)} \mleft( e^{i \frac{2 n_2 \pi}{N} m'} - 1 \mright) + e^{- i \frac{2 n_2 \pi}{N} a} \mleft( e^{- i \frac{2 n_2 \pi}{N} m'} - 1 \mright) \mright] \mright|^2 \ da \nn\\
&=& \frac{\gamma \tau}{2} \sum_{m' = 1}^N \int_{- 1 / 2}^{1 / 2} \mleft|
\frac{A (n_1) e^{- i \frac{2 n_1 \pi}{N \tau} \mleft(t + \frac{\tau}{2} \mright)}}{1 - e^{- i \frac{2 n_1 \pi}{N}}} \mleft[ e^{i \frac{2 n_1 \pi}{N} a} \mleft( e^{i \frac{2 n_1 \pi}{N} m'} - 1 \mright) + e^{- i \frac{2 n_1 \pi}{N} a} \mleft( e^{- i \frac{2 n_1 \pi}{N} m'} - 1 \mright) \mright] \mright. \nn\\
&& \mleft. \qquad \qquad \qquad \qquad
+ \frac{A (n_2) e^{- i \frac{2 n_2 \pi}{N \tau} \mleft( t + \frac{\tau}{2} \mright)}}{1 - e^{- i \frac{2 n_2 \pi}{N}}} \mleft[ e^{i \frac{2 n_2 \pi}{N} a} \mleft( e^{i \frac{2 n_2 \pi}{N} m'} - 1 \mright) + e^{- i \frac{2 n_2 \pi}{N} a} \mleft( e^{- i \frac{2 n_2 \pi}{N} m'} - 1 \mright) \mright] \mright|^2 \ da \nn\\
&=& \frac{\gamma \tau}{2} \sum_{m' = 1}^N \int_{- 1 / 2}^{1 / 2} \mleft|
\frac{2 A (n_1) e^{- i \frac{2 n_1 \pi}{N \tau} \mleft( t + \frac{\tau}{2} \mright)}}{1 - e^{- i \frac{2 n_1 \pi}{N}}} \mleft[ \cos \mleft( \frac{2 n_1 \pi}{N} [m' + a] \mright) - \cos \mleft( \frac{2 n_1 \pi}{N} a \mright) \mright] \mright. \nn\\
&& \mleft. \qquad \qquad \qquad \qquad
+ \frac{2 A (n_2) e^{- i \frac{2 n_2 \pi}{N \tau} \mleft( t + \frac{\tau}{2} \mright)}}{1 - e^{- i \frac{2 n_2 \pi}{N}}} \mleft[ \cos \mleft( \frac{2 n_2 \pi}{N} [m' + a] \mright) - \cos \mleft( \frac{2 n_2 \pi}{N} a \mright) \mright] \mright|^2 \ da \nn\\
&=& \frac{\gamma \tau}{2} \sum_{m' = 1}^N \int_{- 1 / 2}^{1 / 2} \mleft|
\frac{4 A (n_1) e^{- i \frac{2 n_1 \pi}{N \tau} \mleft( t + \frac{\tau}{2} \mright)}}{1 - e^{- i \frac{2 n_1 \pi}{N}}} \sin \mleft( \frac{n_1 \pi}{N} m' \mright) \sin \mleft( \frac{n_1 \pi}{N} [m' + 2 a] \mright) \mright. \nn\\
&& \mleft. \qquad \qquad \qquad \qquad
+ \frac{4 A (n_2) e^{- i \frac{2 n_2 \pi}{N \tau} \mleft( t + \frac{\tau}{2} \mright)}}{1 - e^{- i \frac{2 n_2 \pi}{N}}} \sin \mleft( \frac{n_2 \pi}{N} m' \mright) \sin \mleft( \frac{n_2 \pi}{N} [m' + 2 a] \mright) \mright|^2 \ da \nn\\
&=& I (n_1) + I (n_2) + \frac{\gamma \tau}{2} 16 A (n_1) A (n_2) \mleft[ \frac{e^{- i \frac{2 (n_1 - n_2) \pi}{N \tau} \mleft( t + \frac{\tau}{2} \mright)}}{\mleft( 1 - e^{- i \frac{2 n_1 \pi}{N}} \mright) \mleft( 1 - e^{i \frac{2 n_2 \pi}{N}} \mright)} + \text{h.c.} \mright] \nn\\
&& \times \sum_{m' = 1}^N \sin \mleft( \frac{n_1 \pi}{N} m' \mright) \sin \mleft( \frac{n_2 \pi}{N} m' \mright) \int_{- 1 / 2}^{1 / 2} \sin \mleft( \frac{n_1 \pi}{N} [m' + 2 a] \mright) \sin \mleft( \frac{n_2 \pi}{N} [m' + 2a] \mright) \ da \nn\\
&=& I (n_1) + I (n_2) + \frac{\gamma \tau}{2} 16 A (n_1) A (n_2) \mleft[ \frac{e^{- i \frac{2 (n_1 - n_2) \pi}{N \tau} t}}{4 \sin \mleft( \frac{n_1 \pi}{N} \mright) \sin \mleft( \frac{n_2 \pi}{N} \mright)} + \text{h.c.} \mright] \nn\\
&& \times \frac{1}{2} \sum_{m' = 1}^N \sin \mleft( \frac{n_1 \pi}{N} m' \mright) \sin \mleft( \frac{n_2 \pi}{N} m' \mright) \int_{- 1 / 2}^{1 / 2} \mleft[ \cos \mleft( \frac{[n_1 - n_2] \pi}{N} [m' + 2 a] \mright) - \cos \mleft( \frac{[n_1 + n_2] \pi}{N} [m' + 2 a] \mright) \mright] \ da \nn\\
&=& I (n_1) + I (n_2) + 2 \gamma \tau \frac{A (n_1) A (n_2)}{\sin \mleft( \frac{n_1 \pi}{N} \mright) \sin \mleft( \frac{n_2 \pi}{N} \mright)} \cos \mleft( \frac{2 [n_1 - n_2] \pi}{N \tau} t \mright) \nn\\
&& \times \sum_{m' = 1}^N \sin \mleft( \frac{n_1 \pi}{N} m' \mright) \sin \mleft( \frac{n_2 \pi}{N} m' \mright) \int_{- 1 / 2}^{1 / 2} \mleft[ \cos \mleft( \frac{[n_1 - n_2] \pi}{N} [m' + 2 a] \mright) - \cos \mleft( \frac{[n_1 + n_2] \pi}{N} [m' + 2 a] \mright) \mright] \ da \nn
\eea
\bea
&=& I (n_1) + I (n_2) + 2 \gamma \tau \frac{A (n_1) A (n_2)}{\sin \mleft( \frac{n_1 \pi}{N} \mright) \sin \mleft( \frac{n_2 \pi}{N} \mright)} \cos \mleft( \frac{2 [n_1 - n_2] \pi}{N \tau} t \mright) \nn\\
&& \times \sum_{m' = 1}^N \sin \mleft( \frac{n_1 \pi}{N} m' \mright) \sin \mleft( \frac{n_2 \pi}{N} m' \mright)
\mleft[ \frac{N}{(n_1 - n_2) \pi} \sin \mleft( \frac{[n_1 - n_2] \pi}{N} \mright) \cos \mleft( \frac{[n_1 - n_2] \pi}{N} m' \mright) \mright. \nn\\
&& \mleft. \qquad \qquad \qquad \qquad \qquad \qquad
- \frac{N}{(n_1 + n_2) \pi} \sin \mleft( \frac{[n_1 + n_2] \pi}{N} \mright) \cos \mleft( \frac{[n_1 + n_2] \pi}{N} m' \mright) \mright] \nn\\
&=& I (n_1) + I (n_2) + 2 \gamma \tau \frac{A (n_1) A (n_2)}{\sin \mleft( \frac{n_1 \pi}{N} \mright) \sin \mleft( \frac{n_2 \pi}{N} \mright)} \cos \mleft( \frac{2 [n_1 - n_2] \pi}{N \tau} t \mright) \nn\\
&& \times \frac{1}{2} \sum_{m' = 1}^N \mleft[ \cos \mleft( \frac{[n_1 - n_2] \pi}{N} m' \mright) - \cos \mleft( \frac{[n_1 + n_2] \pi}{N} m' \mright) \mright] \nn\\
&& \times \mleft[ \frac{N}{(n_1 - n_2) \pi} \sin \mleft( \frac{[n_1 - n_2] \pi}{N} \mright) \cos \mleft( \frac{[n_1 - n_2] \pi}{N} m' \mright)
- \frac{N}{(n_1 + n_2) \pi} \sin \mleft( \frac{[n_1 + n_2] \pi}{N} \mright) \cos \mleft( \frac{[n_1 + n_2] \pi}{N} m' \mright) \mright] \nn\\
&=& I (n_1) + I (n_2) + 2 N \gamma \tau \frac{A (n_1) A (n_2)}{4 \sin \mleft( \frac{n_1 \pi}{N} \mright) \sin \mleft( \frac{n_2 \pi}{N} \mright)} \mleft[ \frac{N}{(n_1 - n_2) \pi} \sin \mleft( \frac{[n_1 - n_2] \pi}{N} \mright)
+ \frac{N}{(n_1 + n_2) \pi} \sin \mleft( \frac{[n_1 + n_2]\pi}{N} \mright) \mright] \cos \mleft( \frac{2 [n_1 - n_2] \pi}{N \tau} t \mright). \quad
\end{eqnarray}
Here, $I (n)$ is defined by \eqref{In}. Using the condition in Eq.~(12) in the main text, we finally obtain
\be
\label{PTbidark2}
I (n_1, n_2) = I (n_1) + I (n_2) - 2 A (n_1) A (n_2) \mleft[ 1 + \frac{(n_1 - n_2) \sin \mleft( \frac{n_1 + n_2}{N} \pi \mright)}{(n_1 + n_2) \sin \mleft( \frac{n_1 - n_2}{N} \pi \mright)} \mright] \cos \mleft( \frac{2 [n_1 - n_2] \pi}{N \tau} t \mright).
\ee
For the two dark modes $\Omega_{n_1} = 2 n_1 \pi / (N \tau)$ and $\Omega_{n_2} = 2 n_2 \pi / (N \tau)$, we have
\bea
\label{AE}
\Omega_{n_1} + \Omega_{n_2} &=& \frac{2 (n_1 + n_2) \pi}{N \tau} 
= \Omega \frac{2 (n_1 + n_2) \pi}{2 n_1 \pi - 2 (n_1 - n_2) \pi \frac{\cot \mleft( \frac{n_1 \pi}{N} \mright)}{\cot \mleft( \frac{n_1 \pi}{N} \mright) - \cot \mleft( \frac{n_2 \pi}{N} \mright)}} \nn\\
&=& \Omega \frac{2 (n_1 + n_2) \pi}{2 n_1 \pi + 2 (n_1 - n_2) \pi \frac{\cos \mleft( \frac{n_1 \pi}{N} \mright) \sin \mleft( \frac{n_2 \pi}{N} \mright)}{\sin \mleft( \frac{[n_1 - n_2] \pi}{N} \mright)}} 
= \Omega \frac{2 (n_1 + n_2) \pi}{2 n_1 \pi + (n_1 - n_2) \pi \frac{\sin \mleft( \frac{[n_1 + n_2] \pi}{N} \mright) - \sin \mleft ( \frac{[n_1 - n_2] \pi}{N} \mright)}{\sin \mleft( \frac{[n_1 - n_2] \pi}{N} \mright)}} \nn\\
&=& \Omega \frac{2 (n_1 + n_2) \pi}{(n_1 + n_1) \pi + (n_1 - n_2) \pi \frac{\sin \mleft( \frac{[n_1 + n_2] \pi}{N} \mright)}{\sin \mleft( \frac{[n_1 - n_2]\pi}{N} \mright)}} \nn\\
&=& \frac{2 \Omega}{1 + \frac{(n_1 - n_2) \sin \mleft( \frac{n_1 + n_2}{N} \pi \mright)}{(n_1 + n_2) \sin \mleft( \frac{n_1 - n_2}{N} \pi \mright)}}.
\eea
In the second step, we again used the condition in Eq.~(12) in the main text. The final result is
\bea
I (n_1, n_2) = I (n_1) + I (n_2) - 4 A (n_1) A (n_2) \frac{\Omega }{\Omega_{n_1} + \Omega_{n_2}}\cos \mleft( \mleft[ \Omega_{n_1} - \Omega_{n_2} \mright] t \mright),
\eea
which is exactly Eq.~(14) in the main text.


\end{document}